\documentclass[12pt]{iopart}

\usepackage{graphicx} 
\begin{document}

\title[High Frequency Geodesic Acoustic Modes in Electron Scale Turbulence]{High Frequency Geodesic Acoustic Modes in Electron Scale Turbulence}

\author{Johan Anderson$^1$, Andreas Skyman$^1$, Hans Nordman$^1$, Raghvendra Singh$^2$ and Predhiman Kaw$^2$}

\address{$^1$ Dept. Geo and Space Sciences, Chalmers University of Technology, SE-412 96 G\"{o}teborg, Sweden}
\address{$^2$ Institute for Plasma Research, Bhat, Gandhinagar, Gujarat, India 382428}
\ead{anderson.johan@gmail.com}
\begin{abstract}
In this work the finite $\beta$-effects of an electron branch of the geodesic acoustic mode (el-GAM) driven by electron temperature gradient (ETG) modes is presented. The work is based on a fluid description of the ETG mode retaining non-adiabatic ions and the dispersion relation for el-GAMs driven non-linearly by ETG modes is derived. The ETG growth rate from the fluid model is compared to the results found from gyrokinetic simulations with good agreement. A new saturation mechanism for ETG turbulence through the interaction with el-GAMs is found, resulting in a significantly enhanced ETG turbulence saturation level compared to the mixing length estimate. It is shown that the el-GAM may be stabilized by an increase in finite $\beta$ as well as by increasing non-adiabaticity. The decreased GAM growth rates is due to the inclusion of the Maxwell stress.
\end{abstract}

\maketitle

\section{Introduction}
Experimental investigations has elucidated on the complex dynamics of the low to high (L--H) plasma confinement mode transition. Evidence of interactions between the the turbulence driven $\vec{E} \times \vec{B}$ zonal flow oscillation or Geodesic Acoustic Mode (GAM) \cite{winsor1968, conway2011, mckee2008, diamond2005, miki2007, chak2007, miki2010, hallatschek2012}, turbulence and the mean equilibrium flows during this transition was found. Furthermore, periodic modulation of flow and turbulence level with the characteristic limit cycle oscillation at the GAM frequency was present \cite{conway2011}. The GAMs are weakly damped by Landau resonances and moreover this damping effect is weaker at the edge suggesting that GAMs are more prominent in the region where transport barriers are expected \cite{mckee2008}.

During recent years investigations on coherent structures such as vortices, streamers and zonal flows ($m=n=0$, where $m$ and $n$ are the poloidal and toroidal mode numbers respectively) revealed that they play a critical role in determining the overall transport in magnetically confined plasmas \cite{diamond2005}. Zonal flows impede transport by shear decorrelation \cite{diamond2005, terry2000}, whereas the GAM is the oscillatory counterpart of the zonal flow ($m=n=0$ in the potential perturbation, $m=1$, $n=0$ in the perturbations in density, temperature and parallel velocity) and thus a weaker effect on turbulence is expected \cite{miki2007, chak2007}.

The electron temperature gradient (ETG) mode driven by a combination of electron temperature gradients and field line curvature effects is a likely candidate for driving electron heat transport \cite{liu1971, horton1988, jenko2000, singh2001, singh2001NF, tangri2005}. The ETG driven electron heat transport is determined by short scale fluctuations that do not influence ion heat transport and is largely unaffected by the large scale flows stabilizing ion-temperature-gradient (ITG) modes. The generation of large scale modes such as zonal flows and GAMs is here realized through the Wave Kinetic Equation (WKE) analysis that is based on the coupling of the micro-scale turbulence with the GAM through the WKE under the assumptions that there is a large separation of scales in space and time. ~\cite{diamond2005, smol2000, smol2002, krommes2000, anderson2002, anderson2006, anderson2007} In non-linear gyrokinetic simulations large thermal transport levels, beyond mixing length estimates, have been observed for a long time~\cite{dorland2000, jenko2000, jenko2002, nevins2006, waltz2008, nakata2012}. In particular, in Ref.~\cite{dorland2000} it is found that saturation produced by secondary modes occurs at long wavelengths causing significant streamer dominated transport, suggestive of a high saturation level for the ETG mode such as that found in this paper. 

It is interesting to note that in simulations, damping of the GAM due to coupling to higher $m$ modes has been found ~\cite{sugama2006, miyato2007, sasaki2008}. It is evident that using the wave kinetic equation such damping would inevitably be found by allowing the energy to be distributed among several modes.

It is well known that linear drift wave can be significantly damped by finite $\beta$ effects and it was recently shown that finite $\beta$ can have a significant impact on secondarily generated modes such as zonal flows~\cite{anderson2012}. Although the electron GAM (el-GAM) was previously explored in Refs ~\cite{anderson2012} and ~\cite{chak2012}, vital physics were however neglected. 

In the present work the finite $\beta$-effects are elaborated on, and numerical quantifications of the frequency and growth rate are given. The finite $\beta$-effects are added in an analogous way compared to the recent work on zonal flows in Ref.~\cite{anderson2011, guzdar2008}. In particular the Maxwell stress is included in the generation of the el-GAM. The frequency of the el-Gam is higher compared to the ion GAM by the square root of the ion-to-electron mass ratio ($\Omega_q(electron)/\Omega_q(ion) \approx \sqrt{m_i/m_e}$ where $\Omega_q(electron)$ and $\Omega_q(ion)$ are the real frequencies of the electron and ion GAMs, respectively.). It is found that similar to the linear growth rate the finite $\beta$ effects are stabilizing the GAM using a mode coupling saturation level. Furthermore, increasing the non-adiabaticity parameter ($\Lambda_e$) can decrease the growth rate through a linear contribution. Finally, the effect of enhanced non-linear saturation level found in \cite{anderson2012} on the GAM growth would lead to a correspondingly increased GAM growth rate.

\section{The linear Electron Temperature Gradient Mode}

In this section we will describe the preliminaries of the electron-temperature-gradient (ETG) mode which we consider under the following restrictions on real frequency and wavelength: $\Omega_i \leq \omega \sim \omega_{\star} << \Omega_e$, $k_{\perp} c_i > \omega > k_{\parallel} c_e$. Here $\Omega_j$ are the respective cyclotron frequencies, $\rho_j$ the Larmor radii and $c_j = \sqrt{T_j/m_j}$ the thermal velocities. The diamagnetic frequency is $\omega_{\star} \sim k_{\theta} \rho_e c_e / L_n$, $k_{\perp}$ and $k_{\parallel}$ are the perpendicular and the parallel wave numbers. The ETG model consists of a combination of ion and electron fluid dynamics coupled through quasineutrality, including finite $\beta$-effects \cite{singh2001, tangri2005}.
\subsection{\label{sec:level3} Ion and impurity dynamics}
In this section, we will start by describing the ion fluid dynamics in the ETG mode description. In the limit $\omega > k_{\parallel} c_e$ the ions are stationary along the mean magnetic field $\vec{B}$ (where $\vec{B} = B_0 \hat{e}_{\parallel}$) whereas in the limit $k_{\perp} c_i >> \omega$, $k_{\perp} \rho_i >> 1$ the ions are unmagnetized. In this paper we will use the non-adabatic responses in the limits $\omega < k_{\perp} c_I < k_{\perp} c_i$, where $c_I = \sqrt{\frac{T_I}{m_I}}$ is the impurity thermal velocity, and we assume that $\Omega_i < \omega < \Omega_e$ are fulfilled for the ions and impurities. In the ETG mode description we can utilize the ion and impurity continuity and momentum equations of the form
\begin{eqnarray} \label{eq:1.1}
\frac{\partial n_j}{\partial t} + n_j \nabla \cdot \vec{v}_j & = & 0, \ \ \mbox{and} \\
m_j n_j \frac{\partial \vec{v}_j}{\partial t} + e n_j \nabla \phi + T_j \nabla n_j & = & 0,
\end{eqnarray}
where $j=i$ for ions and $j=I$ for impurities. Now, we derive the non-adiabatic ion response
with $\tau_i = T_e/T_i$ and impurity response with with $\tau_I = T_e/T_I$, respectively. We thus have
\begin{eqnarray}
\widetilde{n}_j = - \left( \frac{z \tau_j}{1 - \omega^2/\left(k_{\perp}^2 c_j^2\right)}\right) \widetilde{\phi}. \label{eq:1.2}
\end{eqnarray}
Here $T_j$ and $n_j$ are the mean temperature and density of species ($j=e,i,I$), where $\widetilde{n}_i = \delta n/ n_i$, $\widetilde{n}_I = \delta n_I/ n_I$ and $\widetilde{\phi} =  e \phi/T_e$ are the normalized ion density, impurity density and potential fluctuations and $z$ is the charge number of species $j$. Next we present the electron dynamics and the linear dispersion relation.
\subsection{\label{sec:level4} The electron model}
The electron dynamics for the toroidal ETG mode are governed by the continuity, parallel momentum and energy equations adapted from the Braginskii fluid equations. The electron equations are analogous to the ion fluid equations used for the toroidal ITG mode,
\begin{eqnarray}
\frac{\partial n_{e}}{\partial t} +\nabla \cdot \left( n_{e} \vec{v}_{E} + n_{e} \vec{v}_{\star e} \right) + \nabla \cdot \left( n_{e} \vec{v}_{pe} + n_{e} \vec{v}_{\pi e} \right) + \nabla \cdot \left(n_e \vec{v}_{\parallel e}\right) & = & 0, \label{eq:1.4} \\ 
 \frac{3}{2} n_{e} \frac{\rmd T_{e}}{\rmd t} + n_{e} T_{e} \nabla \cdot \vec{v}_{e} + \nabla \cdot \vec{q}_{e} & = & 0. \label{eq:1.5}
\end{eqnarray}
Here we used the definitions $\vec{q}_e = - (5 p_e/2m_e \Omega_e) \hat{e}_{\parallel} \times \nabla T_e$ as the diamagnetic heat flux, $\vec{v}_{E}$ is the $\vec{E} \times \vec{B}$ drift, $\vec{v}_{\star e}$ is the electron diamagnetic drift velocity, $\vec{v}_{Pe}$ is the electron polarization drift velocity, $\vec{v}_{\pi}$ is the stress tensor drift velocity, and the derivative is defined as $d\rm/\rmd t = \partial/\partial t + \rho_e c_e \hat{e}_{\parallel} \times \nabla \widetilde{\phi} \cdot \nabla$. A relation between the parallel current density and the parallel component of the vector potential ($A_{\parallel}$) can be found using Amp\`{e}re's law,
\begin{eqnarray}\label{eq:1.6}
\nabla^2_{\perp} \widetilde{A}_{\parallel} = - \frac{4 \pi}{c} \widetilde{J}_{\parallel}.
\end{eqnarray}
Taking into account the diamagnetic cancellations in the continuity and energy equations, the Eqs.~(\ref{eq:1.4}, \ref{eq:1.5} and \ref{eq:1.6}) can be simplified and written in normalized form as
\begin{eqnarray}
- \frac{\partial \widetilde{n}_e}{\partial t}  - \nabla_{\perp}^2 \frac{\partial}{\partial t} \widetilde{\phi} - \left( 1  + \left(1 + \eta_e\right) \nabla_{\perp}^2\right) \frac{1}{r} \frac{\partial}{\partial \theta} \widetilde{\phi} - \nabla_{\parallel} \nabla_{\perp}^2 \widetilde{A}_{\parallel}  & + & \nonumber  \\ \epsilon_n \left( \cos \theta \frac{1}{r}\frac{\partial}{\partial \theta} + \sin \theta \frac{\partial}{\partial r} \right)\left(\widetilde{\phi} - \widetilde{n}_e - \widetilde{T}_e\right) & = & \nonumber \\  
-\left(\beta_e/2\right) \left[\widetilde{A}_{\parallel},\nabla_{\parallel}^2 \widetilde{A}_{\parallel}\right] + \left[ \widetilde{\phi}, \nabla^2 \widetilde{\phi}\right], & & \label{eq:1.101} \\
\left(\left(\beta_e/2 - \nabla_{\perp}^2\right) \frac{\partial}{\partial t} + \left(1+\eta_e\right)\left(\beta_e/2\right)\nabla_y\right)\widetilde{A}_{\parallel} + \nabla_{\parallel} \left(\widetilde{\phi} - \widetilde{n}_e - \widetilde{T}_e\right) & = & \nonumber \\
- \left(\beta_e/2\right) \left[\widetilde{\phi} - \widetilde{n}_e, \widetilde{A}_{\parallel}\right]
+ \left(\beta_e/2\right)\left[\widetilde{T}_e,\widetilde{A}_{\parallel}\right] + \left[\widetilde{\phi}, \nabla_{\perp}^2 \widetilde{A}_{\parallel}\right], \label{eq:1.102} \\
\frac{\partial}{\partial t}\widetilde{T}_e + \frac{5}{3} \epsilon_n \left( \cos \theta \frac{1}{r}\frac{\partial}{\partial \theta} + \sin \theta \frac{\partial}{\partial r} \right) \frac{1}{r}\frac{\partial}{\partial \theta} \widetilde{T}_e + \left(\eta_e - \frac{2}{3} \right) \frac{1}{r}\frac{\partial}{\partial \theta} \widetilde{\phi} - \frac{2}{3} \frac{\partial}{\partial t} \widetilde{n}_e & = & -\left[\widetilde{\phi},\widetilde{T}_e\right]. \label{eq:1.103} \nonumber \\
\end{eqnarray}
Note that similar equations have been used previously in estimating the zonal flow generation in ETG turbulence and have been shown to give good agreement with linear gyrokinetic calculations \cite{singh2001, tangri2005}. The variables are normalized according to
\begin{eqnarray}
\left(\widetilde{\phi}, \widetilde{n}, \widetilde{T}_e\right) & = & \left(L_n/\rho_e\right)\left(e \delta \phi/T_{eo}, \delta n_e/n_0, \delta T_e/T_{e0}\right), \\ \label{eq:1.11}
\widetilde{A}_{\parallel} & = & \left(2 c_e L_n/\beta_e c \rho_e\right) e A_{\parallel}/T_{e0}, \\ \label{eq:1.12}
\beta_e & = &  8 \pi n T_e/B_0^2,  \\ \label{eq:1.13}
\epsilon_n & = & \frac{2 L_n}{R}, \\ \label{eq:1.14}
\eta_e & = & \frac{L_n}{L_{T_e}}. \label{eq:1.15}
\end{eqnarray}
Here, $R$ is the major radius and $[A,B] = \frac{\partial A}{\partial x} \frac{\partial B}{\partial y} - \frac{\partial A}{\partial y} \frac{\partial B}{\partial x}$ is the Poisson bracket. The gradient scale length is defined as $L_f = - (d \ln f/dr)^{-1}$. Using the Poisson equation in combination with (\ref{eq:1.2}) we then find
\begin{eqnarray}
\widetilde{n}_e = - \left( \frac{\tau_i n_i/n_e}{1 - \omega^2/k_{\perp}^2 c_i^2} + \frac{\left(Z^2 n_I/n_e\right) \tau_I}{1 - \omega^2/\left(k_{\perp}^2 c_I^2\right)} + k_{\perp}^2 \lambda_{De}^2\right) \widetilde{\phi}. \label{eq:1.3}
\end{eqnarray}
First we will consider the linear dynamical equations (\ref{eq:1.101}, \ref{eq:1.102} and \ref{eq:1.103}) and utilizing Eq.~(\ref{eq:1.3}) as in Ref.~\cite{tangri2005} and we find a semi-local dispersion relation as follows,
\begin{eqnarray} \label{eq:1.16}
\left[ \omega^2 \left(  \Lambda_e  + \frac{\beta_e }{2} (1 + \Lambda_e ) \right) + \left( 1 - \bar{\epsilon}_n (1 + \Lambda_e) \right) \omega_{\star} \right. & + & \nonumber \\
\left. k_{\perp}^2 \rho_e^2 \left(  \omega - (1 + \eta_e) \omega_{\star} \right) \right] \left( \omega - \frac{5}{3} \bar{\epsilon}_n \omega_{\star} \right) & + & \nonumber \\
\left( \bar{\epsilon}_n \omega_{\star} - \frac{\beta_e}{2} \omega\right) \left( (\eta_e - \frac{2}{3})\omega_{\star} + \frac{2}{3} \omega \Lambda_e \right) & = & \nonumber \\
c_e^2k_{\parallel}^2 k_{\perp}^2 \rho_e^2 \left( \frac{(1 + \Lambda_e) \left( \omega - \frac{5}{3}\bar{\epsilon}_n \omega_{\star}\right) - \left( \eta_e - \frac{2}{3}\right)\omega_{\star} - \frac{2}{3} \omega \Lambda_e}{\omega \left( \frac{\beta_e}{2} + k_{\perp}^2 \rho_e^2\right) - \frac{\beta_e}{2} \left( 1 + \eta_e \right) \omega_{\star}} \right).
\end{eqnarray}
In the following we will use the notation $\Lambda_e = \tau_i (n_i/n_e)/(1 - \omega^2/k_{\perp}^2 c_i^2) + \tau_I (z_\mathrm{eff} n_I/n_e)/(1 - \omega^2/k_{\perp}^2 c_I^2) + k_{\perp}^2 \lambda_{De}^2$. Here we define $z_\mathrm{eff} \approx  z^2 n_I/n_e$. Note that in the limit $T_i = T_e$, $\omega<k_{\perp} c_i$, $k_{\perp} \lambda_{De} < k_{\perp} \rho_e \leq 1$ and in the absence of impurity ions, $\Lambda_e \approx 1$ and the ions follow the Boltzmann relation in the standard ETG mode dynamics. Here $\lambda_{De} = \sqrt{T_c/(4 \pi n_e e^2)}$ is the Debye length, the Debye shielding effect is important for $\lambda_{De}/\rho_e > 1$. The dispersion relation Eq.~(\ref{eq:1.16}) is analogous to the toroidal ion-temperature-gradient mode dispersion relation except that the ion quantities are exchanged to their electron counterparts. Eq.~(\ref{eq:1.16}) is derived by using the ballooning mode transform equations for the wave number and the curvature operator,
\begin{eqnarray}
\nabla_{\perp}^2 \widetilde{f} & = & - k_{\perp}^2 \widetilde{f} = - k_{\theta}^2 \left(1 + \left(s \theta - \alpha \sin \theta\right)^2 \right) \widetilde{f}, \\
\nabla_{\parallel} \widetilde{f} & = & \rmi k_{\parallel} \widetilde{f} \approx \frac{\rmi}{qR} \frac{\partial \widetilde{f}}{\partial \theta}, \\
\widetilde{\epsilon}_n \widetilde{f} & = & \epsilon_n \left( \cos \theta + \left(s \theta - \alpha \sin \theta\right) \sin \theta\right) \widetilde{f}.
\end{eqnarray} 
The geometrical quantities will be determined using a semi-local analysis by assuming an approximate eigenfunction while averaging the geometry dependent quantities along the field line. The form of the eigenfunction is assumed to be
\begin{eqnarray}\label{eq:1.17}
\Psi(\theta) = \frac{1}{\sqrt{3 \pi}}(1 + \cos \theta) \;\;\;\;\; \mbox{with} \;\;\;\;\; |\theta| < \pi.
\end{eqnarray}
In the dispersion relation we will replace $k_{\parallel} = \left< k_{\parallel} \right>$, $k_{\perp} = \left< k_{\perp} \right>$ and $\omega_D = \left< \omega_D \right>$ by the averages defined through the integrals
\begin{eqnarray}
\left< k_{\perp}^2 \right> & = & \frac{1}{N\left(\Psi\right)}\int_{-\pi}^{\pi} d \theta \Psi k_{\perp}^2 \Psi = k_{\theta}^2 \left( 1 + \frac{s^2}{3} \left(\pi^2 - 7.5\right) - \frac{10}{9} s \alpha + \frac{5}{12} \alpha^2 \right), \label{eq:1.18} \\
\left< k_{\parallel}^2 \right> & = & \frac{1}{N\left(\Psi\right)} \int_{-\pi}^{\pi} d \theta \Psi k_{\parallel}^2 \Psi = \frac{1}{3 q^2 R^2}, \label{eq:1.19}\\
\left< \omega_D \right> & = & \frac{1}{N\left(\Psi\right)} \int_{-\pi}^{\pi} d \theta \Psi \omega_D \Psi = \epsilon_n \omega_{\star} \left( \frac{2}{3} + \frac{5}{9}s - \frac{5}{12} \alpha \right) = \epsilon_n g \omega_{\star},  \label{eq:1.20} \\
\left< k_{\parallel} k_{\perp}^2 k_{\parallel} \right> & = & \frac{1}{N\left(\Psi\right)} \int_{-\pi}^{\pi} d \theta \Psi k_{\parallel} k_{\perp}^2 k_{\parallel} \Psi = \frac{k_{\theta}^2}{3 \left(qR\right)^2} \left( 1 + s^2 \left( \frac{\pi^2}{3} - 0.5 \right) - \frac{8}{3}s \alpha + \frac{3}{4} \alpha^2 \right), \label{eq:1.21} \nonumber \\
\\
N(\Psi) & = & \int_{-\pi}^{\pi} d \theta \Psi^2. \label{eq:1.22}
\end{eqnarray}
Here we have from the equilibrium $\alpha = \beta q^2 R \left(1 + \eta_e + (1 + \eta_i) \right)/(2 L_n)$ and $\beta = 8 \pi n_o (T_e + T_i)/B^2$ is the plasma $\beta$, $q$ is the safety factor and $s = r q^{\prime}/q$ is the magnetic shear. The $\alpha$-dependent term above (in Eq.\ref{eq:1.16}) represents the effects of Shafranov shift.

\begin{figure}[ht!]
\centering
\includegraphics[width=7cm, height = 6cm]{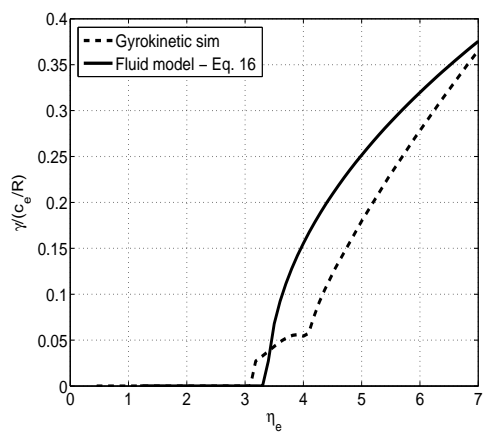}
\caption{The ETG growth rate normalized to $c_e/R$, fluid model (solid line) and gyrokinetic model (dashed line), as a function of $\eta_e$ is shown for the parameters $\epsilon_n = 0.909$, $\beta = 0.01$, $k_{\theta} \rho_e = 0.3$, $s=1$, $\bar{q}=1$ and $\Lambda_e = 2$.}
\label{fig:fig1}
\end{figure}

To test the accuracy of the simple dispersion relation (\ref{eq:1.16}), a comparison with linear, local, gyrokinetic computations of ETG drift wave growth rates has been performed in Figure~\ref{fig:fig1}. The simulations have been carried out with the {\sc GENE} code~\cite{jenko2000} using a simple large aspect ratio, $s-\alpha$ tokamak equilibrium. The simulations were performed in a flux-tube domain at mid-radius ($r/a\approx0.5$) with a resolution of $16\times 24$ grid points in the normal and parallel directions, and $12\times 48$ grid points in momentum space. A quantitatively good agreement between the analytical and numerical results is found. Note that, the linear ETG growth rate and frequency are only weakly dependent on finite $\beta$.

\section{Modeling Electron Geodesic Acoustic modes}
The Geodesic Acoustic Modes are the $m=n=0$, $k_r \neq 0$ perturbation of the potential field and the $n=0$, $m=1$, $k_r \neq 0$ perturbation in the density, temperatures and the magnetic field perturbations.~\cite{winsor1968, diamond2005} The el-GAM ($q, \Omega_q$) induced by ETG modes ($k,\omega$) is considered under the conditions when the ETG mode real frequency satisfies $\Omega_e > \omega > \Omega_i$ at the scale $k_{\perp } \rho_e < 1$ and the real frequency of the GAM fulfils $\Omega_q \sim c_e/R$ at the scale $q_r < k_{r}$. 

\subsection{Linear Electron Geodesic Acoustic Modes}
We start by deriving the linear electron GAM dispersion relation following the outline in the previous paper Ref.~\cite{anderson2012}, by writing the $m=1$ equations for the density, parallel component of the vector potential and temperature, and the $m=0$ of the electrostatic potential
\begin{eqnarray} 
- \tau_i \frac{\partial \widetilde{n}^{(1)}_{eG}}{\partial t} + \epsilon_n \sin \theta \frac{\partial}{\partial r} \widetilde{\phi}^{(0)}_G - \nabla_{\parallel} \nabla_{\perp}^2 \widetilde{A}_{\parallel G}^{(1)} & = & 0, \label{eq:3.101} \\ 
\left(\beta_e/2 - \nabla_{\perp}^2\right) \frac{\partial}{\partial t}\widetilde{A}^{(1)}_{\parallel G} - \nabla_{\parallel} \left( \widetilde{n}^{(1)}_{eG} + \widetilde{T}^{(1)}_{eG} \right) & = & 0, \label{eq:3.102} \\
\frac{\partial}{\partial t}\widetilde{T}^{(1)}_{eG} - \frac{2}{3} \frac{\partial}{\partial t} \widetilde{n}^{(1)}_{eG} & = & 0, \label{eq:3.103} \\
- \nabla_{\perp}^2 \frac{\partial}{\partial t} \widetilde{\phi}^{(0)}_G - \epsilon_n \sin \theta \frac{\partial}{\partial r} \left( \widetilde{n}^{(1)}_{eG} + \widetilde{T}^{(1)}_{eG} \right) & = & 0. \label{eq:3.104}
\end{eqnarray} 
We will derive the linear GAM frequency as follows by using Eq.~(\ref{eq:3.102}) and by eliminating the density $m=1$ component using a time derivative of Eq.~(\ref{eq:3.104}). Finally by utilizing Eq.~(\ref{eq:3.101}) we find,
\begin{eqnarray} \label{eq:3.8}
\rho_e^2 \frac{\partial^2}{\partial t^2} \nabla_{\perp}^2 \widetilde{\phi}^{(0)} +  \epsilon_n v_{\star} \left< \sin \theta \frac{\partial}{\partial r} \left( \epsilon_n v_{\star} \sin \theta \frac{\partial \widetilde{\phi}^{(0)}}{\partial r} + \nabla_{\parallel} \frac{J_{\parallel}^{(1)}}{e n_0} \right) \right> = 0.
\end{eqnarray}
Here $\left< \cdots \right>$ is the average over the poloidal angle $\theta$. In the simplest case this leads to the dispersion relation,
\begin{eqnarray}  \label{eq:3.9}
\Omega_q^2 = \frac{5}{3}\frac{c_e^2}{R^2} \left( 2 + \frac{1}{\bar{q}^2} \frac{1}{1 + \beta_e/\left(2 q_r^2\right)} \right). 
\end{eqnarray}
Here, $\bar{q}$ is the safety factor. Note that the linear electron GAM is purely oscillating analogously to its ion counterpart c.f. Ref.~\cite{chak2007}. Here it is of interest to note that it is very similar to the result found in Ref. ~\cite{chak2012}.

\subsection{\label{sec:level7} The Non-linearly Driven Geodesic Acoustic modes}
We will now study the system including the non-linear terms and derive the electron GAM growth rate. The non-linear extension to the evolution equations presented previously in Eqs.~(\ref{eq:1.101})--(\ref{eq:1.103}) are
\begin{eqnarray} 
- \frac{\partial \widetilde{n}_e}{\partial t}  - \nabla_{\perp}^2 \frac{\partial}{\partial t} \widetilde{\phi} - \left( 1  + \left(1 + \eta_e\right) \nabla_{\perp}^2\right)\nabla_{\theta} \widetilde{\phi} - \nabla_{\parallel} \nabla_{\perp}^2 \widetilde{A}_{\parallel}  & + & \nonumber  \\ \epsilon_n \left( \cos \theta \frac{1}{r}\frac{\partial}{\partial \theta} + \sin \theta \frac{\partial}{\partial r} \right)\left(\widetilde{\phi} - \widetilde{n}_e - \widetilde{T}_e\right) & = &  \nonumber \\ 
+ \left[ \widetilde{\phi}, \nabla^2 \widetilde{\phi}\right] - \left(\beta_e/2\right) \left[\widetilde{A}_{\parallel}, \nabla^2 \widetilde{A}_{\parallel}\right], \label{eq:7.101} \\ 
\left(\left(\beta_e/2 - \nabla_{\perp}^2\right) \frac{\partial}{\partial t} + \left(1+\eta_e\right)\left(\beta_e/2\right)\nabla_{\theta}\right)\widetilde{A}_{\parallel} + \nabla_{\parallel} \left(\widetilde{\phi} - \widetilde{n}_e - \widetilde{T}_e\right) & = & \left[\widetilde{\phi}, \nabla_{\perp}^2 \widetilde{A}_{\parallel}\right], \label{eq:7.102} \nonumber \\ \\
\frac{\partial}{\partial t}\widetilde{T}_e + \frac{5}{3} \epsilon_n \left( \cos \theta \frac{1}{r}\frac{\partial}{\partial \theta} + \sin \theta \frac{\partial}{\partial r} \right) \frac{1}{r}\frac{\partial}{\partial \theta} \widetilde{T}_e + \left( \eta_e - \frac{2}{3} \right) \frac{1}{r}\frac{\partial}{\partial \theta} \widetilde{\phi} - \frac{2}{3} \frac{\partial}{\partial t} \widetilde{n}_e & = & -\left[\widetilde{\phi},\widetilde{T}_e\right]. \label{eq:7.103} \nonumber \\ 
\end{eqnarray}
In order to find the relevant equations for the electron GAM dynamics we consider the $m=1$ component of Eq.~(\ref{eq:7.101}) - (\ref{eq:7.103}),
\begin{eqnarray} 
- \frac{\partial \widetilde{n}^{(1)}_{eG}}{\partial t} + \epsilon_n \sin \theta \frac{\partial}{\partial r} \widetilde{\phi}^{(0)}_G - \nabla_{\parallel} \nabla_{\perp}^2 \widetilde{A}_{\parallel G}^{(1)} & = & \nonumber \\
\left< \left[\widetilde{\phi}_k, \nabla^2 \widetilde{\phi}_k\right] \right>^{(1)} - \left(\beta_e/2\right) \left< \left[\widetilde{A}_{\parallel k}, \nabla^2 \widetilde{A}_{\parallel k}\right] \right>^{(1)} & = & 0, \label{eq:3.111} \\
\left(\beta_e/2 - \nabla_{\perp}^2\right) \frac{\partial}{\partial t}\widetilde{A}^{(1)}_{\parallel G} - \nabla_{\parallel} \left(\widetilde{n}^{(1)}_{eG} + \widetilde{T}^{(1)}_{eG}\right) & = &  \nonumber \\
 \left< \left[\widetilde{\phi}_k, \nabla_{\perp}^2 \widetilde{A}_{\parallel k}\right] \right>^{(1)} & = & 0, \label{eq:3.112}\\
\frac{\partial}{\partial t}\widetilde{T}^{(1)}_{eG} - \frac{2}{3} \frac{\partial}{\partial t} \widetilde{n}^{(1)}_{eG} & = & \nonumber \\
-\left< \left[\widetilde{\phi}_k,\widetilde{T}_{ek}\right]\right>^{(1)} & = & N_1^{(1)}, \label{eq:3.113}
\end{eqnarray}
where superscript (1) over the fluctuating quantities denotes the $m=1$ poloidal mode number, $\left< \cdots \right> $ is the average over the fast time and spatial scales of the ETG turbulence, and that non-linear terms associated with parallel dynamics are small since $\frac{1}{\bar{q}^2} << 1$. Note that, in evaluating the non-linear terms a summation over the spectrum is performed and that the $m=1$ non-linear terms are odd and thus yield a negligible contribution to the non-linear generation of the GAM, assuming a symmetric spectrum, c.f Eq.~(\ref{eq:3.116}). We now study the $m=0$ potential perturbations,
\begin{eqnarray} \label{eq:3.2}
- \nabla_{\perp}^2 \frac{\partial}{\partial t} \widetilde{\phi}^{(0)}_G - \epsilon_n \sin \theta \frac{\partial}{\partial r} \left( \widetilde{n}^{(1)}_{eG} + \widetilde{T}^{(1)}_{eG} \right) & = & \nonumber \\
 \left< \left[\widetilde{\phi}_k, \nabla^2 \widetilde{\phi}_k\right] \right>^{(0)} - \left(\beta_e/2\right) \left< \left[\widetilde{A}_{\parallel k}, \nabla^2 \widetilde{A}_{\parallel k}\right] \right>^{(0)} & = & N_2^{(0)}.
\end{eqnarray}
In order to evaluate the Maxwell stress part in Eq.~(\ref{eq:3.2}), we will approximate the parallel part of the electromagnetic vector potential with the electrostatic potential through a linear relation. The relation $\widetilde{A}_{k \parallel} = A_0 \widetilde{\phi}_k$ is found by using the Eqs.~(\ref{eq:1.102}), (\ref{eq:1.103}) and the non-adiabatic response Eq.~(\ref{eq:1.3}) giving an approximation of the total stress of the form
\begin{eqnarray} \label{eq:3.222}
N_2^{(0)} = (1 - |\Omega_{\alpha}|^2)\left< \left[\widetilde{\phi}_k, \nabla^2 \widetilde{\phi}_k\right] \right>^{(0)}.
\end{eqnarray}
The $\Omega_{\alpha}$ factor is found by using Eq.~(\ref{eq:1.102})
\begin{eqnarray} \label{eq:3.223}
\left|\Omega_{\alpha}\right|^2 = \frac{\beta_e}{2}\left| \frac{k_{\parallel} (1 + \Lambda_e + \varphi_0)}{(\beta_e/2 + k^2_{\perp})\omega - (1 + \eta_e)\beta_e k_{\theta}/2}\right|^2,
\end{eqnarray}
where $\varphi_0$ is determined by the temperature equation
\begin{eqnarray} \label{eq:3.224}
\widetilde{T}_{ek} = \frac{(\eta_e - 2/3)- 2/3\omega \Lambda_e}{\omega + 5/3 \epsilon_n g k_{\theta}} \widetilde{\phi}_k = \varphi_0 \widetilde{\phi}_k,
\end{eqnarray}
and $\Lambda_e$ is determined by the non-adiabatic response condition. The expression Eq. (~\ref{eq:3.223}) for the magnetic flutter non-linearity is comparable to that found in Ref.~\cite{singh2001NF} except that in Eq. (~\ref{eq:3.223}) the adiabatic response is taken into account. Note that $\Omega_{\alpha}$ vanishes at $\beta_e = 0$. In the above we defined non-linear term on the RHS in Eqs.~(\ref{eq:3.111})--(\ref{eq:3.2}) as an average over the fast time and small spatial scales of the ETG turbulence such that only small scale self interactions are important. This can be written $\widetilde{T}_e = \frac{2}{3} \widetilde{n}_e^{(1)} + \int dt N_2^{1}$, where the $m=1$ component is determined by an integral of the convective non-linear term. This leads to a relation between the $m=1$ component of the density and temperature fluctuations modified by a non-linear term. Here, the non-linear terms can be written in the form
\begin{eqnarray} 
N_1^{(1)} & = &  \overline{\left(\partial_x \widetilde{\phi}_k \partial_y \widetilde{T}_{ek}\right)} \nonumber \\
& = & \sum_k k_{\theta}^2 \frac{\eta_e \gamma}{\left|\omega\right|^2} \nabla_r \left|\widetilde{\phi}_k\right|^2, \label{eq:3.116} \\
N_2^{(0)} & = & \left(\nabla_x^2 - \nabla_y^2\right) \left( \overline{\left(\partial_x \widetilde{\phi}_k \partial_y \widetilde{\phi}_k\right)} +  \frac{\beta_e}{2}\overline{\left(\partial_x \widetilde{A}_{\parallel k} \partial_y \widetilde{A}_{\parallel k}\right)} \right) \nonumber \\
& = & \left(1 - \left|\Omega_{\alpha}\right|^2\right) q_r^2 \sum_k k_r k_{\theta} \left|\widetilde{\phi}_k \right|^2. \label{eq:3.117}
\end{eqnarray}

We continue by considering the Eqs.~(\ref{eq:3.111}) and~(\ref{eq:3.2}) for the $m=1$ component and $m=0$ component, respectively,
\begin{eqnarray} 
\frac{\partial \widetilde{n}^{(1)}_{eG}}{\partial t} - \frac{\nabla_{\parallel} \widetilde{J}_{\parallel}^{(1)}}{e n_0} - \epsilon_n \sin \theta \frac{\partial \widetilde{\phi}_G^{(0)}}{\partial r} = N_1^{(1)}, \label{eq:3.71}\\
\frac{\partial}{\partial t} \nabla_{\perp}^2 \widetilde{\phi}_G^{(0)} +  \epsilon_n \left< \sin \theta \frac{\partial}{\partial r} \left( \frac{5}{3} \widetilde{n}_{eG}^{(1)} + N_1^{(1)}\right) \right> = N_2^{(0)}. \label{eq:3.72}
\end{eqnarray}
Here, we keep the $N_1^{(1)}$ non-linear term in order to quantify the effects of the convective non-linearity and similarly to the operations performed to find the linear electron GAM frequency we eliminate the $m=1$ component of the electron density by taking a time derivative of Eq.~(\ref{eq:3.72}) and use Eq.~(\ref{eq:3.71}) to subsitute the time derivative of the GAM density. This yields
\begin{eqnarray} \label{eq:3.10}
\frac{\partial^2}{\partial t^2} \nabla_{\perp}^2 \widetilde{\phi}_G^{(0)} +  \epsilon_n \left< \sin \theta \frac{\partial}{\partial r} \left( \frac{5}{3}\left( \epsilon_n \sin \theta \frac{\partial \widetilde{\phi}_G^{(0)}}{\partial r} +  \nabla_{\parallel} \frac{J_{\parallel}^{(1)}}{e n_0} + N_1^{(1)} \right) + \frac{\partial}{\partial t} N_1^{(1)} \right) \right> = \frac{\partial}{\partial t}N_2^{(0)}. \nonumber \\
\end{eqnarray}
Note that the el-GAM wave equation will be modified by the effects of the parallel current density ($\widetilde{J}_{\parallel}$) and the $m=1$ non-linear terms in the general case, however we see by inspection that on average the term $N_1^{(1)}$ does not contribute whereas the $N_2^{(0)}$ non-linearity may drive the GAM unstable. We will use the wave kinetic equation \cite{diamond2005, smol2000, smol2002, krommes2000, chak2007, anderson2002, anderson2006, anderson2007} to describe the background short scale ETG turbulence for $(\Omega_q, q) < (\omega, k)$, where the action density $N_k = E_k/|\omega_r| \approx \epsilon_0 |\phi_k|^2/\omega_r$. Here $\epsilon_0 |\phi_k|^2$, is the total energy in the ETG mode with mode number $k$ where $\epsilon_0 = \tau + k_{\perp}^2 + \frac{\eta_e^2 k_\theta^2}{|\omega|^2}$. We note that for typical edge plasmas, where the density profiles are peaked, $L_n/R \sim 0.05$, and the turbulent scale length could be larger with $k_{\theta} \rho_e \sim 0.5$, the scale separation requirement is still fulfilled. The electrostatic potential is represented as a sum of fluctuating and mean quantities $ \label{eq:4.1} \phi(\vec{X},\vec{x},T,t) = \Phi(\vec{X},T) + \widetilde{\phi}(\vec{x},t)$, where $\Phi(\vec{X},T)$ is the mean flow potential. The coordinates $\left( \vec{X}, T\right)$, $\left( \vec{x},t \right)$ are the space and time coordinates for the mean flows and small scale fluctuations, respectively. The wave kinetic equation can be written as
\begin{eqnarray} \label{eq:4.2}
\frac{\partial }{\partial t} N_k(r,t) & + & \frac{\partial }{\partial k_r} \left( \omega_k + \vec{k} \cdot \vec{v}_g \right)\frac{\partial N_k(r,t)}{\partial r} - \frac{\partial }{\partial r} \left( \vec{k} \cdot\vec{v}_g\right) \frac{\partial N_k(r,t)}{\partial k_r} \nonumber \\
& = &  \gamma_k N_k(r,t) - \Delta\omega N_k(r,t)^2.
\end{eqnarray}
We will solve Equation (\ref{eq:4.2}) by assuming a small perturbation ($\delta N_k$) driven by a slow variation for the GAM compared to the mean ($N_{k0}$) such that $N_k = N_{k 0} + \delta N_k$. The relevant non-linear terms can be approximated in the following form
\begin{eqnarray} 
\left< \left[\widetilde{\phi}_k, \nabla_{\perp}^2 \widetilde{\phi}_k\right] \right> & \approx & \left(1 - \left|\Omega_{\alpha}\right|^2\right) q_r^2 \sum_k k_r k_{\theta} \frac{\left|\omega_r\right|}{\epsilon_0} \delta N_k\left(\vec{q},\Omega_q\right). \label{eq:4.3}
\end{eqnarray}
For all GAMs we have $q_r > q_{\theta}$, with the following relation between $\delta N_k$ and $\partial N_{k 0} / \partial k_r$,
\begin{eqnarray} \label{eq:4.6}
\delta N_k = -\rmi q_r^2 k_{\theta} \phi^0_G G(\Omega_q) \frac{\partial N_{0k}}{\partial k_r} + \frac{k_{\theta} q_r \widetilde{T}_{e G}^{(1)} N_{0k}}{\tau_i \sqrt{\eta_e - \eta_{eth}}}, 
\end{eqnarray}
where we have used $\delta \omega_q = \vec{k} \cdot \vec{v}_{Eq}\approx \rmi (k_{\theta} q_r - k_r q_{\theta}) \phi^{(0)}_G$ in the wave kinetic equation and the definition $G(\Omega_q) = \frac{1}{\Omega_q - q_r v_{gr} + \rmi \gamma_k}$. Here the linear instability threshold of the ETG mode is denoted by $\eta_{e th}$ and is determined by numerically solving Eq.~(\ref{eq:1.16}).
Using the results from the wave-kinetic treatment we can compute the non-linear contributions to be of the form
\begin{eqnarray}
\left< \left[\widetilde{\phi}, \nabla_{\perp}^2 \widetilde{\phi}\right] \right> & = & \nonumber \\
& -\rmi& \left(1 - |\Omega_{\alpha}|^2\right) q_r^4 \sum k_r k_{\theta}^2 \frac{|\omega_r|}{\epsilon_0} G\left(\Omega_q\right) \frac{\partial N_k}{\partial k_r} \widetilde{\phi}_G^{(0)} \nonumber \\
& + & \left(1 - |\Omega_{\alpha}|^2\right) \frac{2}{3} q_r^3 \sum k_r k_{\theta} \frac{|\omega_r|}{\epsilon_0} \frac{R N_0}{\tau_i \sqrt{\eta_e - \eta_{th e}}} \widetilde{n}_{e G}^{(1)}. \label{eq:4.61}
\end{eqnarray}
In order to find the non-linear growth rate of the electron GAM we need to find relations between the variables $\widetilde{n}_{e G}^{(1)}$, $\widetilde{T}_{e G}^{(1)}$ and $\widetilde{\phi}_G^{(0)}$,
\begin{eqnarray}
\widetilde{n}_{e G}^{(1)} & = & - \frac{\epsilon_n q_r \sin \theta  \Omega_q}{\Omega_q^2 - \frac{5}{3}\frac{q_{\parallel}^2 q_r^2}{ q_r^2 + \beta_e/2}} \widetilde{\phi}_G^{(0)}. \label{eq:4.8}
\end{eqnarray}
Using Eq.~(\ref{eq:4.8}) and the Fourier representation of Eq.~(\ref{eq:3.10}) resulting in
\begin{eqnarray} \label{eq:4.11}
& & q_r^2 \Omega_q^3 - \frac{5}{3} \frac{q_{\parallel}^2 q_r^4}{\beta_e/2 + q_r^2} \Omega_q - \frac{5}{6} \epsilon_n^2 q_r^2 \Omega_q  =  \nonumber \\
& & \left(1 - |\Omega_{\alpha}|^2\right) \left(\Omega_q^2 - \frac{5}{3}\frac{q_{\parallel}^2 q_r^2}{\beta_e/2 + q_r^2} \right) q_r^4 \sum k_r k_{\theta}^2 \frac{|\omega_r|}{\epsilon_0} G(\Omega_q) \frac{\partial N_k}{\partial k_r}.
\end{eqnarray}
Eq.~(\ref{eq:4.11}) is the sought dispersion relation for the electron GAM. Note that Eq.~\ref{eq:4.11}, reduces to Eq.~(64) of Ref. (\cite{anderson2012}) provided that the spectrum is symmetric. In the electrostatic limit we find a perturbative solution of the form,
\begin{eqnarray}
\frac{\gamma_q}{c_e/R} & \approx & \frac{1}{2} \frac{q_r^2 \rho_e^2 k_{\theta} \rho_e}{\sqrt{\epsilon_n (\eta_e)}} \frac{1}{1 + 1/2\bar{q}^2} \left| \widetilde{\phi}_k \frac{L_n}{\rho_e}\right|^2. \label{eq:4.14}
\end{eqnarray}
Here the main contribution to the non-linear generation of el-GAMs originates from the Reynolds stress term in competition with the Maxwell stress term. Note that the Maxwell stress term reduces the drive of the GAM analogously to the zonal flow situation ~\cite{anderson2011} and that this result differs from the result found in ~\cite{anderson2012} where $\Omega_{\alpha} = 0$. The non-linearly driven electron GAM is unstable with a growth rate depending on the saturation level $\left| \widetilde{\phi}_k \right|^2$ of the ETG mode turbulence. Eq.~\ref{eq:4.14}, gives an estimate of the maximum growth rate where a monochromatic wave packet is assumed for the wave action density.
\begin{figure}[ht!]
\centering
\includegraphics[width=7cm, height = 6cm]{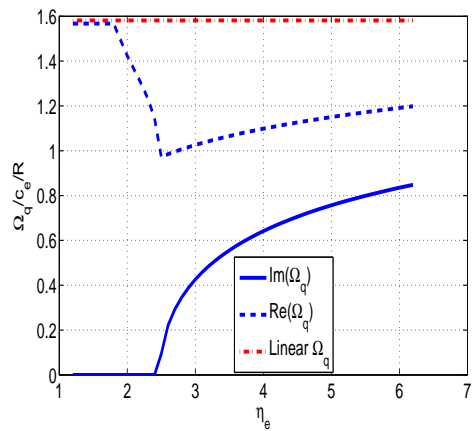}
\caption{(color online) The el-GAM growth (blue, solid line) normalized ($c_e/R$), real frequency (blue, dashed line), the linear solution (red, dash-dotted line) and for the non-linear saturation level (black dotted line) of Eq. (\ref{eq:5.2}) as a function of $\eta_e$ is shown for the parameters $\epsilon_n = 0.909$, $\beta = 0.01$, $k_{\theta} \rho_e = 0.6$, $k_{\parallel} = 0.01$, $s = 1$ and $\bar{q}=1$ in the strong ballooning limit $g(\theta)=1$ with ETG saturation level $|\widetilde{\phi}_k| \approx (\gamma_k/\omega_{\star}) (1 / k_r L_n)$.}
\label{fig:fig2}
\end{figure}

\begin{figure}[ht!]
\centering
\includegraphics[width=7cm, height = 6cm]{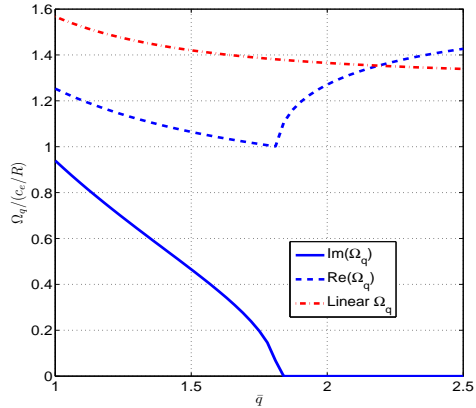}
\caption{(color online) The el-GAM growth (blue, solid line) normalized to ($c_/R$), real frequency (blue, dashed line) and the linear solution (red, dash-dotted line) as a function of the safety factor $\bar{q}$ is shown for the parameter $\eta_e = 4.0$ whereas the remaining parameters are as in Figure 2.}
\label{fig:fig3}
\end{figure}

We will now present the numerically determined frequencies and growth rates found by solving Eq.~(\ref{eq:4.11}) in a few limits. We have chosen common parameters relevant for an ETG mode situation with $\eta_e = 4.0$, $\epsilon_n = 0.909$, $\beta = 0.01$ $k_{\theta} \rho_e = 0.6$ and $\bar{q}=1$ in the strong ballooning limit $g(\theta)=1$. Furthermore the saturation level of the ETG mode is represented by the mode coupling estimate $|\widetilde{\phi}_k| \approx (\gamma_k/\omega_{\star}) (1 / k_r L_n)$. This is motivated by non-linear simulations of ITG turbulence~\cite{nordman1989}, however we will show that an elevated saturation state is available. Moreover, of the possible saturation states, we have used the mode coupling saturation level for simplicity whereas the enhanced saturation level found in Eq. (\ref{eq:5.2}) would lead to an corresponding increase in the GAM growth rate.

In Figure 2, we display the el-GAM ($\Omega_q$) growth rate and real frequency (in blue) as a function of $\eta_e$ in comparison with the solution to the linear el-GAM dispersion relation Eq.~(\ref{eq:3.9} in red) in the strong ballooning limit of $g(\theta)=1$. The el-GAM growth rate is increasing with increasing $\eta_e$, analogously to the linear ETG growth rate due to the mode coupling saturation level which is proportional to the linear ETg growth rate. Note that the threshold value of the GAM instability is dependent of the linear instability threshold of the ETG mode given by $\eta_{e th}$.

Figure 3 shows the el-GAM ($\Omega_q$) growth rate and real frequency (in blue) as a function of $\bar{q}$ in comparison with the solution to the linear el-GAM dispersion relation Eq.~(\ref{eq:3.9} in red) in the strong ballooning limit of $g(\theta)=1$. It is found that $\bar{q}$ has a stabilizing effect on the el-GAM. However, the quantitative results are strongly dependent on the parameters while the finite $\beta$-effects are moderate. Here, the effect of the Maxwell stress on the growth rate is rather small of the order of $\beta_e \sim 0.01$ for the given parameters, however, for increasing amount of impurity ions (increasing $z_\mathrm{eff}$) it becomes increasingly important. Note that, in the GAM dispersion relation a threshold value for the non-linear drive is present below which the el-GAM is stable.   

In Figure 4, we expound on the effects of finite $\beta$ and the non-adiabaticity $\Lambda_e$. The el-GAM growth rate ($\Im (\Omega_q)$) (solid lines) and frequency ($\Re (\Omega_q)$) (dashed lines) as a function of $\beta$ with $\Lambda_e = 1.0$ (blue) and $\Lambda_e = 2.0$ (red) as a parameter are shown. We find that, similar to the linear growth rate, the el-GAM is stablized by the finite $\beta$ effects. Furthermore increasing $\Lambda_e$ can further decrease the growth rate through a linear contribution.  

\begin{figure}[ht!]
\centering
\includegraphics[width=7cm, height = 6cm]{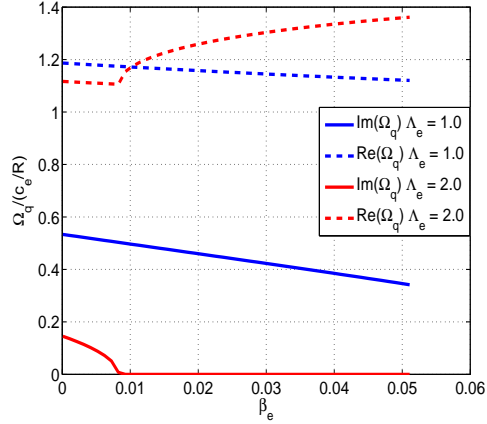}
\caption{The el-GAM growth rates (solid line) and real frequencies (dashed lines) normalized to $c_e/R$ as a function of $\beta_e$ with the non-adiabaticity {$\Lambda_e$} as a parameter. The remaining parameters are as in Figure 2 for the parameters with $s=1$, $q=1$ in the semi-local analysis.}
\label{fig:fig4}
\end{figure}

\section{Saturation mechanism}
Large thermal transport levels, beyond mixing length estimates, have been observed in gyrokinetic simulations for a long time~\cite{dorland2000, jenko2000, jenko2002, nevins2006, waltz2008, nakata2012}. In particular, in Ref.~\cite{dorland2000} it is found that saturation produced by secondary modes occurs at long wavelengths causing significant streamer dominated transport. Moreover, Ref.\cite{waltz2008} highlights the importance of the GAM for the saturation and that the GAM frequency is involved in the non-linear saturation process since at large $\bar{q}$ the GAM growth rate may vanish. In this section we will estimate a new saturation level to provide an alternative elevated saturation level for the ETG turbulence seen in experiments and simulations. The saturation level for the ETG turbulent electrostatic potential ($\widetilde{\phi}_k$) is estimated by balancing the Landau damping in competition with the non-linear growth rate of the GAM in a constant background of ETG mode turbulence, according to the well known predator-prey models used, c.f. Eq.~(4) in Ref.~\cite{miki2007}, 
\begin{eqnarray} \label{eq:5.1}
\frac{\partial N_k}{\partial t} & = & \gamma_k N_k - \Delta \omega N_k^2 - \gamma_1 U_G N_k, \\
\frac{\partial U_G}{\partial t} & = & \gamma_q U_G - \gamma_L U_G - \nu^{\star} U_G.
\end{eqnarray}
Here we have represented the ETG mode turbulence as $N_k = |\phi_k|^2 \frac{L_n^2}{\rho_e^2}$ and $U_G = \left< \frac{ e \phi_G^{(0)}}{T_e} \frac{L_n}{\rho_e} \sin \theta \right>$ with the following parameters: $\gamma$ is the ETG mode growth rate, $\gamma_1$ is the coupling between the ETG mode and the GAM. The Landau damping rate $\left( \gamma_L = \frac{4 \sqrt{2}}{3 \sqrt{\pi}} \frac{c_e}{\bar{q}R} \right)$ is assumed to be balanced by GAM growth rate Eq.~(\ref{eq:4.14}) modified by the neoclassical damping in stationary state $ \frac{\partial N}{\partial t} \rightarrow 0$ and $ \frac{\partial U_G}{\partial t} \rightarrow 0$. In steady state we find the saturation level for the ETG turbulent intensity as ($\gamma_q = \gamma_L + \nu^{\star}$)
\begin{eqnarray} \label{eq:5.2}
\left|\frac{e \phi_k}{T_e} \frac{L_n}{\rho_e} \right|^2 \approx \frac{2 L_n}{\bar{q} R} \left(1 + \frac{1}{2 \bar{q}^2} \right) \sqrt{\epsilon_n \eta_e} \left(\frac{4}{3} \sqrt{\frac{2}{\pi}} + \nu^{\star}\right) \left(\frac{k_{\theta}}{q_r} \right)^2 \left( \frac{1}{k_{\theta} \rho_e} \right)^3.
\end{eqnarray}
Here, the saturation level is modified by the neoclassical damping $\nu^{\star} = \nu_{e} \frac{\bar{q}R}{v_{eth}}$ and the $\frac{k_{\theta}}{q_r}$ factor arises due to the spatial extension of the GAM and we obtain
\begin{eqnarray} \label{eq:5.3}
\left|\frac{e \phi_k}{T_e} \frac{L_n}{\rho_e} \right| \sim 30 - 40.
\end{eqnarray}
Note that this is significantly larger than the mixing length estimate with $\left| \frac{e \phi}{T_e} \frac{L_n}{\rho_e}\right| \sim 1$. In this estimation we have used values of the parameters relevant for an experimental edge plasma such that $L_n = 0.05$, $\bar{q} = 3.0$, $R = 4$, $\epsilon_n = 0.025$, $1/q_r \sim (\rho_e^2 L_T)^{1/3}$, $k_{\theta} \rho_e = 0.3$ where $k_{\theta} / q_r \approx 4$ and $\eta_e  \sim 1$.

\section{Conclusions}
In this work the electromagnetic effects on the electron Geodesic Acoustic Mode (el-GAM) are investigated. The work extends a previous study (Ref.~\cite{anderson2012}) by self-consistently including linear as well as non-linear $\beta$ effects in the derivation. The linear dispersion relation of the el-GAM is purely oscillatory with a frequency $\Omega_q \sim \frac{c_e}{R}$ whereas the GAM growth rate is estimated by a non-linear treatment based on the wave-kinetic approach. The el-GAM growth rate is driven by a competition between the Reynolds stress and the Maxwell stress. The non-linear dispersion relation is solved numerically where it is found that the magnetic safety factor $\bar{q}$ has a stabilizing effect on the el-GAM. However the quantitative results are strongly dependent on the other physical parameters while the finite $\beta$-effects are moderate. The effect of the Maxwell stress on the el-GAM growth rate can be significant and the GAM can be stabilized for increasing $\beta$. Moreover, for an increasing amount of impurity ions (increasing $z_\mathrm{eff}$), the $\beta$ effects become increasingly important. Note that, in the dispersion relation a set-off non-linear drive is present below which the el-GAM is stable. 

To estimate the ETG mode fluctuation level and GAM growth, a predator-prey model was used to describe the coupling between the GAMs and small scale ETG turbulence. The stationary point of the coupled system implies that the ETG turbulent saturation level $\widetilde{\phi}_k$ can be drastically enhanced by the new saturation mechanism, stemming from a balance between the Landau damping and the GAM growth rate. This may result in highly elevated particle and electron heat transport, relevant for the edge pedestal region of H-mode plasmas.

\end{document}